\documentclass[12pt, showkeys, showpacs ] {revtex4}
\usepackage{graphicx}
\begin{document}
\title[Short Title]{Probabilistic Dense Coding Using a Non-symmetric Multipartite Quantum Channel}

\author{Qiu-Bo \surname{Fan}}
\author{Shou \surname{Zhang}\footnote{E-mail: szhang@ybu.edu.cn}}
\affiliation{Department of Physics, College of Science, Yanbian
University, Yanji, Jilin, 133002, PR China}

\begin{abstract}  We investigate probabilistic dense coding
in non-symmetric Hilbert spaces of the sender's and the receiver's
particles. The sender and the receiver share the multipartite
non-maximally quantum channel. We also discuss the average
information.
\keywords{Probabilistic dense coding; Non-symmetric
Hilbert space; Quantum channel}
 \pacs{03.67.-a, 89.70.+c}
\end{abstract}
\maketitle Quantum dense coding \cite{01} is one of the important
applications of quantum entangled state \cite{02,03} in quantum
communication. Some people have proposed some schemes
\cite{04,05,06,07,08} for quantum dense coding using mixed state
entanglement, Greenberger-Horne-Zeilinger (GHZ) state, multi-level
entangled state,and multipartite entangled state. Other people
have proposed a number of schemes about dense coding \cite{09,10}
using the interaction between atoms and cavity quantum
electrodynamics (QED). However, the quantum channels of these
schemes are all symmetrically and maximally entangled states. And
then Yan {\it et al}. \cite{11} and Fan {\it et al}. \cite{12}
have discussed dense coding using two-particle and multi-particle
entangled states as quantum channels, respectively. In their
schemes the quantum channels are all non-symmetric and maximal.
Wang {\it et al}. \cite{13} have proposed another scheme for
probabilistic dense coding using a non-maximally and symmetrically
entangled pair. And Pati {\it et al}. \cite{14} have also proposed
a scheme for probabilistic super dense coding with non-maximally
and symmetrically entangled states as a resource, and they
generalized the scheme to higher dimension and more entanglement.
There are other people proposed distributed quantum dense coding
\cite{15}, i.e., the generalization of quantum dense coding to
more than one sender and more than one receiver. But in our
scheme,we investigate probabilistic dense coding using two
non-symmetrically and non-maximally entangled pairs as quantum
channel and generalize it to $N$ non-symmetrically and
non-maximally entangled pairs, that it to say, our quantum channel
is a non-symmetric and multi-particle state. Our scheme only has
one sender and one receiver.

Now we discuss our scheme in detail. For clarity, we first use
four particles $1, 2, 1'$, and $2'$ to realize the probabilistic
dense coding. We suppose particles 1 and 2, both in 3-dimension
Hilbert space, belong to Alice; particles $1'$ and $2'$, both in
2-dimension Hilbert space, belong to Bob. The initial state which
they share is as follows:
\begin{equation}\label{e1}
|\Psi\rangle_{121'2'}=(\alpha_{0_{1}}|00\rangle+\alpha_{1_{1}}|11\rangle)_{11'}\otimes(\alpha_{0_{2}}|00\rangle
+\alpha_{1_{2}}|11\rangle)_{22'},
\end{equation}
where the subscripts 1 and 2 of $\alpha_{0}$ and $\alpha_{1}$
indicate the 1, 2-th two-particle entangled state, respectively;
$\alpha_{0_{1}}, \alpha_{1_{1}}, \alpha_{0_{2}}$, and
$\alpha_{1_{2}}$ are real numbers; and
$|\alpha_{0_{1}}|^{2}+|\alpha_{1_{1}}|^{2}=1,
|\alpha_{0_{2}}|^{2}+|\alpha_{1_{2}}|^{2}=1 $. Without loss of
generality, we suppose that $|\alpha_{0_{1}}|\leq
|\alpha_{1_{1}}|$ and $|\alpha_{0_{2}}|\leq |\alpha_{1_{2}}|$.

Firstly, Alice introduces two auxiliary two-level particles in the
quantum state $|00\rangle_{a_{1}a_{2}}$. So the total state of the
system is
\begin{equation}\label{e2}
|\Psi\rangle_{\rm
T}=(\alpha_{0_{1}}|00\rangle+\alpha_{1_{1}}|11\rangle)_{11'}\otimes(\alpha_{0_{2}}|00\rangle+\alpha_{1_{2}}|11\rangle)_{22'}\otimes|00\rangle_{a_{1}a_{2}}.
\end{equation}
Alice performs two unitary operations $U_{1a_{1}}$ and
$U_{2a_{2}}$ on her particles $(1, a_{1})$ and $(2, a_{2})$, under
the basis $\{|00\rangle, |01\rangle, |10\rangle, |11\rangle,
|20\rangle, |21\rangle\}_{1a_{1}(2a_{2})}$,  respectively. We can
write the two unitary operations as one $U=U_{1a_{1}}\otimes
U_{2a_{2}}$ as follows:
\begin{equation}\label{e3}
U=U_{1a_{1}}\otimes U_{2a_{2}}=\left[%
\begin{array}{cccccc}
  1\ &\ 0 &\ 0 &\ 0 &\ 0 &\ 0\\
  0\ &\ 1 &\ 0 &\ 0 &\ 0 &\ 0\\
  0\ &\ 0 &\ A_{1} &\ B_{1} &\ 0 &\ 0\\
  0\ &\ 0 &\ B_{1} &\ -A_{1} &\ 0 &\ 0\\
  0\ &\ 0 &\ 0 &\ 0 &\ 1 &\ 0\\
  0\ &\ 0 &\ 0 &\ 0 &\ 0 &\ 1\\
\end{array}%
\right]\otimes\left[%
\begin{array}{cccccc}
  1\ &\ 0 &\ 0 &\ 0 &\ 0 &\ 0\\
  0\ &\ 1 &\ 0 &\ 0 &\ 0 &\ 0\\
  0\ &\ 0 &\ A_{2} &\ B_{2} &\ 0 &\ 0\\
  0\ &\ 0 &\ B_{2} &\ -A_{2} &\ 0 &\ 0\\
  0\ &\ 0 &\ 0 &\ 0 &\ 1 &\ 0\\
  0\ &\ 0 &\ 0 &\ 0 &\ 0 &\ 1\\
\end{array}%
\right],
\end{equation}
where $A_{i}=\alpha_{0_{i}}/\alpha_{1_{i}},
B_{i}=\sqrt{(1-\alpha_{0_{i}}^{2}/\alpha_{1_{i}}^{2})}(i=1,2)$.
The unitary operation will transform $|\Psi\rangle_{\rm T}$ into
the corresponding state:
\begin{eqnarray}\label{e4}
|\Psi'\rangle_{\rm
T}&=&\left\{\alpha_{0_{1}}(|00\rangle+|11\rangle)_{11'}|0\rangle_{a_{1}}+\sqrt{\alpha_{1_{1}}^{2}-\alpha_{0_{1}}^{2}}|11\rangle_{11'}|1\rangle_{a_{1}}\right\}
\cr\cr&&\otimes\left\{\alpha_{0_{2}}(|00\rangle+|11\rangle)_{22'}|0\rangle_{a_{2}}+\sqrt{\alpha_{1_{2}}^{2}-\alpha_{0_{2}}^{2}}|11\rangle_{22'}|1\rangle_{a_{2}}\right\}
\cr\cr&=&\left\{\sqrt{2}\alpha_{0_{1}}\frac{1}{\sqrt{2}}(|00\rangle+|11\rangle)_{11'}|0\rangle_{a_{1}}+\sqrt{\alpha_{1_{1}}^{2}-\alpha_{0_{1}}^{2}}|11\rangle_{11'}|1\rangle_{a_{1}}\right\}
\cr\cr&&\otimes\left\{\sqrt{2}\alpha_{0_{2}}\frac{1}{\sqrt{2}}(|00\rangle+|11\rangle)_{22'}|0\rangle_{a_{2}}+\sqrt{\alpha_{1_{2}}^{2}-\alpha_{0_{2}}^{2}}|11\rangle_{22'}|1\rangle_{a_{2}}\right\}
\cr\cr&=&\left\{\sqrt{2}\alpha_{0_{1}}|\Psi_{00}^{0}\rangle_{11'}|0\rangle_{a_{1}}+\sqrt{\alpha_{1_{1}}^{2}-\alpha_{0_{1}}^{2}}|\Psi_{00}^{1}\rangle_{11'}|1\rangle_{a_{1}}\right\}
\cr\cr&&\otimes\left\{\sqrt{2}\alpha_{0_{2}}|\Psi_{00}^{0}\rangle_{22'}|0\rangle_{a_{2}}+\sqrt{\alpha_{1_{2}}^{2}-\alpha_{0_{2}}^{2}}|\Psi_{00}^{1}\rangle_{22'}|1\rangle_{a_{2}}\right\}.
\end{eqnarray}
where
$|\Psi_{00}^{0}\rangle_{11'(22')}=\frac{1}{\sqrt{2}}(|00\rangle+|11\rangle)_{11'(22')},
|\Psi_{00}^{1}\rangle_{11'(22')}=|11\rangle_{11'(22')}$. Then
Alice makes orthogonal measurement on the auxiliary particles. If
she gets the result $|0\rangle_{a_{1}}|0\rangle_{a_{2}}$, she
ensures that the four particles $1$, $1'$, $2$, and $2'$ are in
the product state of the two maximally entangled pairs, i.e.,
$|\Psi_{00}^{0}\rangle_{11'}\otimes|\Psi_{00}^{0}\rangle_{22'}$,
and the probability of obtaining
$|0\rangle_{a_{1}}|0\rangle_{a_{2}}$ is
$4\alpha_{0_{1}}^{2}\alpha_{0_{2}}^{2}$ according to
Eq.~(\ref{e4}); if she gets the result
$|0\rangle_{a_{1}}|1\rangle_{a_{2}}$, she ensures that the four
particles are in the state
$|\Psi_{00}^{0}\rangle_{11'}\otimes|\Psi_{00}^{1}\rangle_{22'}$,
and the probability of this result is
$2\alpha_{0_{1}}^{2}(\alpha_{1_{2}}^{2}-\alpha_{0_{2}}^{2})$; if
she gets the result $|1\rangle_{a_{1}}|0\rangle_{a_{2}}$, she
ensures that the four particles are in the state
$|\Psi_{00}^{1}\rangle_{11'}\otimes|\Psi_{00}^{0}\rangle_{22'}$,
and the probability is
$2\alpha_{0_{2}}^{2}(\alpha_{1_{1}}^{2}-\alpha_{0_{1}}^{2})$; if
she gets the result $|1\rangle_{a_{1}}|1\rangle_{a_{2}}$, she
ensures the four particles are in the state
$|\Psi_{00}^{1}\rangle_{11'}\otimes|\Psi_{00}^{1}\rangle_{22'}$,
and the probability is
$(\alpha_{1_{1}}^{2}-\alpha_{0_{1}}^{2})(\alpha_{1_{2}}^{2}-\alpha_{0_{2}}^{2})$.

Secondly, Alice encodes classical information by a unitary
transformation on her particles 1 and 2. If particles 1 and $1'$(2
and $2'$) are in the state
$|\Psi_{00}^{0}\rangle_{11'(22')}=\frac{1}{\sqrt{2}}(|00\rangle+|11\rangle)_{11'(22')}
$, she can perform six single-particle operators on particle 1(2):
\begin{equation}\label{e5}
\begin{array}{cc}
U_{00}^{0}=\left[%
\begin{array}{ccc}
  1\ &\ 0\ &\ 0 \\
  0\ &\ 1\ &\ 0 \\
  0\ &\ 0\ &\ 1 \\
\end{array}%
\right],
U_{01}^{0}=\left[%
\begin{array}{ccc}
  1 & 0 & 0 \\
  0 & -1 & 0 \\
  0 & 0 & 1 \\
\end{array}%
\right],
U_{10}^{0}=\left[%
\begin{array}{ccc}
  0\ &\ 0\ &\ 1 \\
  1\ &\ 0\ &\ 0 \\
  0\ &\ 1\ &\ 0 \\
\end{array}
\right],
\cr\\
U_{11}^{0}=\left[%
\begin{array}{ccc}
  0 & 0 & 1 \\
  1 & 0 & 0 \\
  0 & -1 & 0 \\
\end{array}%
\right],
U_{20}^{0}=\left[%
\begin{array}{ccc}
  0\ &\ 1\ &\ 0 \\
  0\ &\ 0\ &\ 1 \\
  1\ &\ 0\ &\ 0 \\
\end{array}%
\right],
U_{21}^{0}=\left[%
\begin{array}{ccc}
  0 & -1 & 0 \\
  0 & 0 & 1 \\
  1 & 0 & 0 \\
\end{array}%
\right].
\end{array}
\end{equation}

The state $|\Psi_{00}^{0}\rangle_{11'(22')}$ will be transformed
into the corresponding states:
\begin{equation}\label{e6}
\begin{array}{cccccc}
U_{00}^{0}\
|\Psi_{00}^{0}\rangle_{11'(22')}=\displaystyle\frac{1}{\sqrt{2}}\
(|00\rangle+|11\rangle)_{11'(22')}=|\Psi_{00}^{0}\rangle_{11'(22')},\cr\\
 U_{01}^{0}\ |\Psi_{00}^{0}\rangle_{11'(22')}=\displaystyle\frac{1}{\sqrt{2}}\
(|00\rangle-|11\rangle)_{11'(22')}=|\Psi_{01}^{0}\rangle_{11'(22')},\cr\\
U_{10}^{0}\
|\Psi_{00}^{0}\rangle_{11'(22')}=\displaystyle\frac{1}{\sqrt{2}}\
(|10\rangle+|21\rangle)_{11'(22')}=|\Psi_{10}^{0}\rangle_{11'(22')},\cr\\
 U_{11}^{0}\ |\Psi_{00}^{0}\rangle_{11'(22')}=\displaystyle\frac{1}{\sqrt{2}}\
(|10\rangle-|21\rangle)_{11'(22')}=|\Psi_{11}^{0}\rangle_{11'(22')},\cr\\
 U_{20}^{0}\ |\Psi_{00}^{0}\rangle_{11'(22')}=\displaystyle\frac{1}{\sqrt{2}}\
(|20\rangle+|01\rangle)_{11'(22')}=|\Psi_{20}^{0}\rangle_{11'(22')},\cr\\
U_{21^{0}}\
|\Psi_{00}^{0}\rangle_{11'(22')}=\displaystyle\frac{1}{\sqrt{2}}\
(|20\rangle-|01\rangle)_{11'(22')}=|\Psi_{21}^{0}\rangle_{11'(22')}.
\end{array}
\end{equation}
The above states are orthogonal mutually.

If particles 1 and $1'$(2 and $2'$) are in the product state
$|\Psi_{00}^{1}\rangle_{11'(22')}=|11\rangle_{11'(22')}$, Alice
can perform three single-particle operators on particle 1(2):
\begin{equation}\label{e7}
\begin{array}{c}
U_{00}^{1}=\left[%
\begin{array}{ccc}
  1\ &\ 0\ &\ 0 \\
  0\ &\ 1\ &\ 0 \\
  0\ &\ 0\ &\ 1 \\
\end{array}%
\right],
U_{10}^{1}=\left[%
\begin{array}{ccc}
  0\ &\ 1\ &\ 0 \\
  1\ &\ 0\ &\ 0 \\
  0\ &\ 0\ &\ 1 \\
\end{array}%
\right],
U_{20}^{1}=\left[%
\begin{array}{ccc}
  1\ &\ 0\ &\ 0 \\
  0\ &\ 0\ &\ 1 \\
  0\ &\ 1\ &\ 0 \\
\end{array}
\right].
\end{array}
\end{equation}
The state $|\Psi_{00}^{1}\rangle_{11'(22')}$ will be transformed
into the corresponding states:
\begin{equation}\label{e8}
\begin{array}{ccc}
U_{00}^{1}\
|\Psi_{00}^{1}\rangle_{11'(22')}=|11\rangle_{11'(22')}=|\Psi_{00}^{1}\rangle_{11'(22')},\cr\\
U_{10}^{1}\
|\Psi_{00}^{1}\rangle_{11'(22')}=|01\rangle_{11'(22')}=|\Psi_{10}^{1}\rangle_{11'(22')},\cr\\
U_{20}^{1}\
|\Psi_{00}^{1}\rangle_{11'(22')}=|21\rangle_{11'(22')}=|\Psi_{20}^{1}\rangle_{11'(22')}.
\end{array}
\end{equation}
These states are also orthogonal mutually.

In all, there are four cases Alice can encode classical information
on her particles 1 and 2 according to the measurement results of the
auxiliary particles $a_{1}$ and $a_{2}$, which have different
probabilities. TABLE I shows the four cases(where the superscripts 1
and 2 of $U^{0}_{m_{0}n_{0}}$, $U^{1}_{m_{1}n_{1}}$ indicate the 1,
2-th particles; the subscripts 0 and 1 of each $\Psi$ and $U$
correspond to the results of each auxiliary particle; $m_{0},
m_{1}=0, 1, 2$; $n_{0}=0, 1$; $n_{1}=0$).
\begin{table}\label{ta1}
\caption{The four cases which are separated by the measurement
results about the auxiliary particles $a_{1}$ and $a_{2}$  .}
{\small
\begin{tabular}{||c|c|c|c|c||}\hline\hline
The results of pa- & The states which & The unitary
operato- & The corresponding & The number\\
rticles $a_{1}$ and $a_{2}$ & Alice ensures  &rs which Alice makes
&  states & of the states\\
\hline
$|0\rangle_{a_{1}}|0\rangle_{a_{2}}$&$|\Psi_{00}^{0}\rangle_{11'}\otimes|\Psi_{00}^{0}\rangle_{22'}$&$(U^{0}_{m_{0}n_{0}})^{1}\otimes
(U^{0}_{m_{0}n_{0}})^{2}$&$|\Psi_{m_{0}n_{0}}^{0}\rangle_{11'}\otimes|\Psi_{m_{0}n_{0}}^{0} \rangle_{22'}$&36\\
\hline
$|0\rangle_{a_{1}}|1\rangle_{a_{2}}$&$|\Psi_{00}^{0}\rangle_{11'}\otimes|\Psi_{00}^{1}\rangle_{22'}$&$(U^{0}_{m_{0}n_{0}})^{1}\otimes
(U^{1}_{m_{1}n_{1}})^{2}$&$|\Psi_{m_{0}n_{0}}^{0}\rangle_{11'}\otimes|\Psi_{m_{1}n_{1}}^{1}\rangle_{22'}$&18\\
\hline
$|1\rangle_{a_{1}}|0\rangle_{a_{2}}$&$|\Psi_{00}^{1}\rangle_{11'}\otimes|\Psi_{00}^{0}\rangle_{22'}$&$(U^{1}_{m_{1}n_{1}})^{1}\otimes
(U^{0}_{m_{0}n_{0}})^{2}$&$|\Psi_{m_{1}n_{1}}^{1}\rangle_{11'}\otimes|\Psi_{m_{0}n_{0}}^{0}\rangle_{22'}$&18\\
\hline
$|1\rangle_{a_{1}}|1\rangle_{a_{2}}$&$|\Psi_{00}^{1}\rangle_{11'}\otimes|\Psi_{00}^{1}\rangle_{22'}$&$(U^{1}_{m_{1}n_{1}})^{1}\otimes
(U^{1}_{m_{1}n_{1}})^{2}$&$|\Psi_{m_{1}n_{1}}^{1}\rangle_{11'}\otimes|\Psi_{m_{1}n_{1}}^{1} \rangle_{22'}$&9\\
\hline\hline
\end{tabular} }
\end{table}

Thirdly, after performing one of these unitary operators on her
particles 1 and 2, Alice sends her particles to Bob and tells Bob
her measurement result of particles $a_{1}$ and $a_{2}$.

Finally, Bob receives the particles, and makes measurements on the
four particles 1, $1'$, 2, and $2'$. The measurement basis is
selected according to Alice's measurement result of particles
$a_{1}$ and $a_{2}$. After that, Bob will obtain the classical
information that Alice has encoded.

From the above procedure, we can calculate the average information
transformation:
\begin{eqnarray}\label{e9}
I_{\rm ave} &=&4\alpha_{0_{1}}^{2}\alpha_{0_{2}}^{2}\log_{2}36
+2\alpha_{0_{1}}^{2}(\alpha_{1_{2}}^{2}-\alpha_{0_{2}}^{2})\log_{2}18\cr\cr
&&+2\alpha_{0_{2}}^{2}(\alpha_{1_{1}}^{2}-\alpha_{0_{1}}^{2})\log_{2}18
+(\alpha_{1_{1}}^{2}-\alpha_{0_{1}}^{2})(\alpha_{1_{2}}^{2}-\alpha_{0_{2}}^{2})\log_{2}9.
\end{eqnarray}
In addition, the above scheme needs $\log_{2}4$ bits of classical
information for Alice to tell Bob her measurement result on the
two auxiliary particles.

We can also generalize the above scheme to arbitrarily different
dimensions Hilbert space for $N$ non-maximally entangled pairs.
Alice and Bob need to share $N$ non-maximally entangled pairs. One
particle of each entangled pair in $p$ dimension belongs to Alice,
and the other in $q$ dimension belongs to Bob (that it to say,
particles 1, 2, $\cdots$, $N$ belong to Alice, and particles $1'$,
$2'$, $\cdots$, $N'$ belong to Bob), where $p\neq q$, i.e., the
Hilbert space of Alice's particles is non-symmetric with that of
Bob's particles. Without loss of generality, we choose $p>q$. The
total state which Alice and Bob share is
\begin{eqnarray}\label{e10}
|\Psi\rangle_{\rm
T}=\bigotimes_{k=1}^{N}(\alpha_{0_{k}}|00\rangle+\alpha_{1_{k}}|11\rangle+\cdots+\alpha_{(q-1)_{k}}|q-1q-1\rangle)_{kk'},
\end{eqnarray}
where $\alpha_{0_{k}},\alpha_{1_{k}},\cdots,\alpha_{(q-1)_{k}}$
are real numbers and satisfy $|\alpha_{0_{k}}|\leq
|\alpha_{1_{k}}|\leq\cdots\leq|\alpha_{(q-1)_{k}}|$.

Similarly, the scheme of probabilistic dense coding can be
realized by the following steps.

Firstly, Alice introduces $N$ auxiliary $q$-level particles in the
quantum state $\bigotimes_{k=1}^{N}|0\rangle_{a_{k}}$. Then she
performs a proper unitary transformation on her particles and the
auxiliary particles. The unitary transformation
$U=\bigotimes_{k=1}^{N}U_{ka_{k}}$ transforms the state
$|\Psi\rangle_{\rm T}\bigotimes_{k=1}^{N}|0\rangle_{a_{k}}$ into
the state
\begin{eqnarray}\label{e11}
|\Psi'\rangle_{\rm
T}&=\bigotimes\limits_{k=1}^{N}&\left\{\alpha_{0_{k}}(|00\rangle+|11\rangle+\cdots+|q-1q-1\rangle)_{kk'}|0\rangle_{a_{k}}\right.\cr\cr
&&+\sqrt{\alpha_{1_{k}}^{2}-\alpha_{0_{k}}^{2}}(|11\rangle+\cdots+|q-1q-1\rangle)_{kk'}|1\rangle_{a_{k}}\cr\cr
&&+\cdots\cr\cr
&&\left.+\sqrt{\alpha_{(q-1)_{k}}^{2}-\alpha_{(q-2)_{k}}^{2}}|q-1q-1\rangle)_{kk'}|q-1\rangle_{a_{k}}\right\}\cr\cr
&=\bigotimes\limits_{k=1}^{N}&\left\{\sqrt{q}\alpha_{0_{k}}\frac{1}{\sqrt{q}}(|00\rangle+|11\rangle+\cdots+|q-1q-1\rangle)_{kk'}|0\rangle_{a_{k}}\right.\cr\cr
&&+\sqrt{q-1}\sqrt{\alpha_{1_{k}}^{2}-\alpha_{0_{k}}^{2}}\frac{1}{\sqrt{q-1}}(|11\rangle+\cdots+|q-1q-1\rangle)_{kk'}|1\rangle_{a_{k}}\cr\cr
&&+\cdots\cr\cr
&&\left.+\sqrt{\alpha_{(q-1)_{k}}^{2}-\alpha_{(q-2)_{k}}^{2}}|q-1q-1\rangle)_{kk'}|q-1\rangle_{a_{k}}\right\}.
\end{eqnarray}
Then Alice makes orthogonal measurement on the auxiliary
particles. She ensures that the quantum channel will be in the
states
$\bigotimes_{k=1}^{N}\frac{1}{\sqrt{q}}(|00\rangle+|11\rangle+\cdots+|q-1q-1\rangle)_{kk'}$,
$\bigotimes_{k=1}^{N-1}\frac{1}{\sqrt{q}}(|00\rangle+|11\rangle+\cdots+|q-1q-1\rangle)_{kk'}\frac{1}{\sqrt{q-1}}(|11\rangle+\cdots+|q-1q-1\rangle)_{NN'}
$, $\cdots$, or $\bigotimes_{k=1}^{N}|q-1q-1\rangle_{kk'}$
corresponding  to the results of auxiliary particles
$\bigotimes_{k=1}^{N}|0\rangle_{a_{k}}$,
$\bigotimes_{k=1}^{N-1}|0\rangle_{a_{k}}|1\rangle_{a_{N}}$,
$\cdots$, or $\bigotimes_{k=1}^{N}|q-1\rangle_{a_{k}}$. The
corresponding probabilities of the results of auxiliary particles
are $q^{N}\prod_{k=1}^{N}\alpha_{0_{k}}^{2}$,
$q^{N-1}(q-1)\prod_{k=1}^{N-1}\alpha_{0_{k}}^{2}(\alpha_{1_{N}}^{2}-\alpha_{0_{N}}^{2})$,
$\cdots$,
$\prod_{k=1}^{N}(\alpha_{(q-1)_{k}}^{2}-\alpha_{(q-2)_{k}}^{2})$.

Secondly, Alice encodes classical information by making a unitary
transformation on her particles $1, 2, \cdots$, and $N$. According
to the results of the auxiliary particles
$\bigotimes_{k=1}^{N}|0\rangle_{a_{k}}$,
$\bigotimes_{k=1}^{N-1}|0\rangle_{a_{k}}|1\rangle_{a_{N}}$,
$\cdots$, or $\bigotimes_{k=1}^{N}|q-1\rangle_{a_{k}}$, the
unitary operations which Alice can make on her particles are
$\bigotimes_{k=1}^{N}(U_{m_{0}n_{0}}^{0})^{k}$,
$\bigotimes_{k=1}^{N-1}(U_{m_{0}n_{0}}^{0})^{k}(U^{1}_{m_{1}n_{1}})^{N}$,
$\cdots$, $\bigotimes_{k=1}^{N}(U^{q-1}_{m_{q-1}n_{q-1}})^{k}$.
The corresponding states are
$\bigotimes_{k=1}^{N}|\Psi_{m_{0}n_{0}}^{0}\rangle_{kk'}$,
$\bigotimes_{k=1}^{N-1}|\Psi_{m_{0}n_{0}}^{0}\rangle_{kk'}|\Psi_{m_{1}n_{1}}^{1}\rangle_{NN'}$,
$\cdots$,
$\bigotimes_{k=1}^{N}|\Psi_{m_{q-1}n_{q-1}}^{q-1}\rangle_{kk'}$.
The unitary operations $U_{m_{0}n_{0}}^{0}$, $U^{1}_{m_{1}n_{1}}$,
$\cdots$, $U^{q-1}_{m_{q-1}n_{q-1}}$ are showed as follows:
\begin{eqnarray}\label{e12}
&&U_{m_{0}n_{0}}^{0}=\sum\limits_{j=0}^{q-1}e^{2\pi
ijn_{0}/q}|(j\oplus m_{0}){\rm mod} p \rangle\langle j|,\cr\cr
&&U^{1}_{m_{1}n_{1}}=\sum\limits_{j=0}^{q-1}e^{2\pi
ijn_{1}/q}|(j\oplus m_{1}){\rm mod}p\rangle\langle j|,\cr\cr
&&\cdots\cr\cr
&&U^{q-1}_{m_{q-1}n_{q-1}}=\sum\limits_{j=0}^{q-1}e^{2\pi
ijn_{q-1}/q}|(j\oplus m_{q-1}){\rm mod}p\rangle\langle j|;
\end{eqnarray}
the states
$|\Psi_{m_{0}n_{0}}^{0}\rangle$,$|\Psi_{m_{1}n_{1}}^{1}\rangle$,
$\cdots$, $|\Psi_{m_{q-1}n_{q-1}}^{q-1}\rangle$ are
\begin{eqnarray}\label{13}
&&|\Psi_{m_{0}n_{0}}^{0}\rangle=\sum\limits_{j=0}^{q-1}e^{2\pi
ijn_{0}/q}|(j\oplus m_{0}) {\rm
mod}p\rangle_{1}\otimes|j\rangle/\sqrt{q},\cr\cr
&&|\Psi_{m_{1}n_{1}}^{1}\rangle=\sum\limits_{j=1}^{q-1}e^{2\pi
ijn_{1}/q}|(j\oplus m_{1}) {\rm
mod}p\rangle_{1}\otimes|j\rangle/\sqrt{q-1},\cr\cr &&\cdots\cr\cr
&&|\Psi_{m_{q-1}n_{q-1}}^{q-1}\rangle=\sum\limits_{j=q-1}^{q-1}e^{2\pi
ijn_{q-1}/q}|(j\oplus m_{q-1}){\rm
mod}p\rangle_{1}\otimes|j\rangle;
\end{eqnarray}
where $m_{0}, m_{1}, \cdots, m_{q-1}=0, 1, \cdots, p-1$; $n_{0}=0,
1, \cdots, q-1$; $n_{1}=0, 1, \cdots, q-2$; $n_{3}=0, 1, \cdots,
q-3$; $\cdots$ ; $n_{q-1}=0$.

Thirdly, Alice sends her particles to Bob, and tells him her
measurement result of the auxiliary particles.

Finally, Bob receives Alice's particles $1, 2, \cdots$, and $N$,
and makes measurement on all his particles according to Alice's
measurement result on the auxiliary particles. Then Bob can obtain
the classical information that Alice has encoded on her particles
via his measurement.

Obviously, the average information Bob can obtain is
\begin{eqnarray}\label{e14}
I_{\rm
ave}&=&q^{N}\prod^{N}_{k=1}\alpha^{2}_{0_{k}}\log_{2}\left[%
(p\times
q)^{N}%
\right]
+q^{N-1}(q-1)\prod_{k=1}^{N-1}\alpha_{0_{k}}^{2}(\alpha_{1_{N}}^{2}-\alpha_{0_{N}}^{2})\log_{2}\left[%
(p\times
q)^{N-1}p(q-1)%
\right]\cr\cr
&&+q^{N-1}(q-2)\prod_{k=1}^{N-1}\alpha_{0_{k}}^{2}(\alpha_{2_{N}}^{2}-\alpha_{1_{N}}^{2})\log_{2}\left[%
(p\times
q)^{N-1}p(q-2)%
\right]\cr\cr
&&+\cdots+\prod_{k=1}^{N}(\alpha_{(q-1)_{k}}^{2}-\alpha_{(q-2)_{k}}^{2})\log_{2}\left[%
p^{N}%
\right].
\end{eqnarray}

In addition, this scheme consumes $N\log_{2}q$ bits of classical
information to transmit Alice's measurement results of the
auxiliary particles.

We discuss the average information transformation $I_{\rm ave}$.
When the quantum channel is composed of two entangled pairs, we can
draw a figure about $\alpha_{0_{1}}, \alpha_{0_{2}}$ and $I_{\rm
ave}$. So from FIG. 1 we can see $I_{\rm ave}$ clearly. When
$\alpha_{0_{1}}^{2}=0.5$ and $ \alpha_{0_{2}}^{2}=0.5$, $I_{\rm
ave}$ is more than 5, i.e., when the two entangled pairs are in the
maximally entangled states, $I_{\rm ave}$ has the maximal value
$\log_{2}36$, more than 5 bits information. Similarly, when we
generalize this scheme to $N$ entangled pairs, if all the entangled
pairs are in maximally entangled states, $I_{\rm {ave}}$ has the
maximal value $N\log_{2}(p\times q)$.
\begin{figure}\label{t1}
\includegraphics{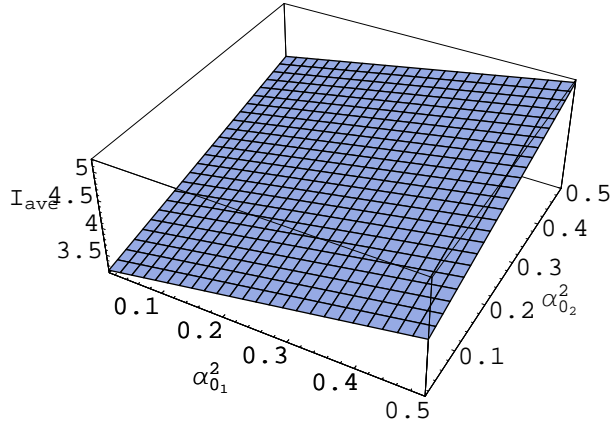}
\caption{The relationship of $\alpha_{0_{1}}, \alpha_{0_{2}}$ and
$I_{\rm ave}$}
\end{figure}

In this scheme, we discuss probabilistic dense coding via two
non-symmetrically and non-maximally entangled pairs as quantum
channel and generalize it to $N$ non-symmetrically and
non-maximally entangled pairs. We also consider the maximal value
of the average information transformation. The average information
has maximal value when all entangled pairs are in the maximally
entangled states.

Comparing with Ref \cite{12} (not probabilistic and with
multipartite quantum channel) which operates with the same
physical resource, our scheme is more general, because the
maximally entangled state is not only difficult for permanently
because of decoherence, but also is hard to be prepared in the
experiment. Therefore, we chose the non-maximally entangled state
as quantum channel. A comparison of the symmetric quantum channel
\cite{13} with our non-symmetric multipartite quantum channel
shows our scheme can increase the efficiency of information
transmission. Comparing with Ref \cite{14}, we use different
quantum channel and different path to realize probabilistic dense
coding. In their scheme, the quantum channel is composed of two
particles and in the same dimensions, and they performed Positive
Operator Valued Measurements (POVMs) on the qubit states to
distinguish these non-orthogonal states. They find that the
success probability of performing super dense coding is exactly
the same as the success probability of distinguishing a set of
non-orthogonal. But in our scheme, we use non-symmetric
multipartite state as quantum channel. The non-symmetric
multipartite state can be converted into orthogonal states by
introducing a set of auxiliary particles and making some unitary
operations. And these orthogonal states can be distinguished only
by some simple measurements. The probability depends on the
measurement results of auxiliary particles. In conclusion, we have
proposed a general efficiency scheme for dense coding.

\end{document}